Use ion beam techniques to study the coupling between air and its relative humidity on iron corrosion under irradiation

S. Lapuerta[a,b], N. Moncoffre[a], N. Bérerd, H. Jaffrezic, N. Millard-Pinard[a], D. Crusset[b]

[a] Institut de Physique Nucléaire de Lyon, 4, rue Enrico Fermi, 69622 Villeurbanne cedex, France,

[b] ANDRA, Parc de la Croix Blanche 1-7 rue Jean Monnet, F-92298 Châtenay-Malabry Cedex, France

Abstract

In this paper, the role of air humidity on the iron corrosion under irradiation is studied in the context of geological disposal of nuclear wastes. The irradiation experiments are performed at room temperature using a 3 MeV extracted proton beam with a 10 nA intensity. Different atmospheres are studied: Humid air with a relative humidity (RH) fixed at 45 %, dry air and a $^{15}N_2$ atmosphere (45% RH). The hydrogen and oxygen distribution profiles at the iron surface in contact with atmosphere are measured by using respectively ERDA (Elastic Recoil Detection Analysis) and RBS (Rutherford Backscattering Spectrometry) analysis. From these experiments it is clearly demonstrated that the coupling of $O_2+H_2O$ enhances iron oxidation whereas for iron hydrogenation, humidity is sufficient whatever the atmosphere. An interpretation is given, which is based on the reaction mechanisms and the species formed by air ionisation.

1. Introduction

In this paper, the role of atmosphere humidity on the iron corrosion under irradiation is studied in the context of geological disposal of high level nuclear wastes. Indeed, in France it is proposed that these wastes will be embedded in stainless steel containers surrounded by carbon steel overpacks. In a first step, the overpacks will be submitted to atmospheric corrosion under γ irradiation. We propose a fundamental study of the iron corrosion under proton irradiation focussed on the role of the surrounding atmosphere. The irradiation are performed with a 3 MeV proton beam characterised mainly by electronic interactions. The resulting ionisations are similar to those created by γ irradiations , but much more intense and concentrated around the protons tracks. Three types of controlled atmosphere have been successively fixed : humid air with a Relative Humidity (RH) of 45%, humid nitrogen atmosphere (RH=45%) and dry air. The aim of this study is to understand the mechanisms responsible for the oxygen and hydrogen gains at the iron surface. Ion beam analysis techniques are used to characterise the evolution of light elements (H, O, N) at the iron − atmosphere interface.

2. Experimental

    2.1 Irradiation conditions

The irradiation experiments are performed using the 4 MV Van de Graaff accelerator of the Nuclear Physics Institute of Lyon (IPNL) which delivers a 3 MeV proton beam. The proton beam is extracted from the beam line vacuum to the atmosphere by crossing a 5 µm thick havar window. The external proton beam already described in [1], enters the irradiation cell through the studied iron foil and stops in water after a 100 µm range. The irradiation line is equipped with a sweeping system which allows a homogeneous irradiation over the whole iron surface (5x5 mm). During irradiation, the beam intensity was kept constant and measured carefully with a calibrated beam chopper. In the experiments presented in this paper, the beam intensity is set to 10 nA .The experimental set up is displayed in Fig. 1. The studied material is a 10 µm thick iron foil (99.85% purity). This foil could not be polished because its weak thickness. As a consequence, the iron surface presents a slight oxide contamination due to the iron foil manufacturing. In the following, it will be referred to as the initial sample. The relative atmosphere humidity in the 8 mm thick gap between the havar window and the iron foil is chosen and measured. The relative humidity control set up is presented in figure 2. The dry gas supply is provided from a bottle whose flow is regulated with a manometer. The dry gas goes first through a water container, and then through an alumina ($Al_2O_3$) trap. The gas flow and the alumina surface in contact with the gas are adjusted in order to reach the right RH equilibrium. Throughout the irradiation experiment the relative humidity is measured using a Hygropalm humidity controller.

The irradiations have been systematically performed at room temperature during 45 minutes. so that only the atmospheric media varies from an irradiation to another. It must be noted that the nitrogen gas was 99% $^{15}N$ and that, in case of irradiation in dry atmosphere, the dry gas flow was introduced directly in the gap.

### 2.2 Ion beam analysis.

Ion beam analysis is performed using the 4 MV Van de Graaff accelerator of the IPNL. RBS and ERDA analysis are used to determine the oxygen and hydrogen profile distributions at the iron surface. RBS is performed at a 172° detection angle and with 1.7 MeV α particles. ERDA is also induced by 1.7 MeV α particles. The incident angle on the target is 15°, the detection angle is 30°. A 6.5 μm thick mylar absorber is placed in front of the Si detector to stop backscattered α particles. In such conditions, the depth resolution is close to 20 nm in iron. A silicium carbide hydrogenated (25 at. % H) sample was used as a normalisation standard. Figure 3 displays the evolution in RBS and ERDA spectra between an initial iron sample and a corroded one. The SIMNRA program is used to simulate the energy spectra obtained both by RBS and ERDA so as to determine the atomic concentration profiles of oxygen and hydrogen. For a given sample analysed by RBS and ERDA, the analysis is done by iteration, taking always into account, in the simulations files, the iron, oxygen and hydrogen concentrations. Finally, the $^{15}$N(p,αγ)$^{12}$C nuclear reaction at 897 keV is used to check that even in a 99% $^{15}$N atmosphere, the nitrogen concentration at the iron surface is extremely low (figure 4). It demonstrates that nitrogen does not play any role on the iron corrosion.

3. Study of the iron corrosion as a function of the atmospheric conditions

Oxygen and hydrogen profiles are respectively presented in figures 5 and 6 as a function of the surrounding medium during irradiation. Let us just mention that the hydrogen contamination due to proton irradiation is negligible, more especially as the proton beam crosses the iron foil.

Concerning the oxygen profiles, figure 5 clearly displays that the oxidation is strongly enhanced up to 60 at.% only for the humid air atmosphere. In the other conditions, the oxygen amount remains the same as for the initial sample. Hoerlé et al. [2] who have studied iron atmospheric corrosion at room temperature have put in evidence the major corrosion during wet cycles in comparison to the dry ones. They explain such a corrosion process by oxydo-reduction reactions:

$$2\ Fe \rightarrow 2\ Fe^{2+} + 4\ e^-$$

$$O_2 + 4\ e^- + 2\ H_2O \rightarrow 4\ OH^-$$

———————————————

$$2\ Fe + O_2 + 2\ H_2O \leftrightarrow 2\ Fe^{2+} + 4\ OH^-$$

These reactions put forward the simultaneous need of $O_2$ and $H_2O$. In addition, Graedel et al. [3] have measured a corrosion rate in carbon steel submitted to air corrosion equal to 104 μm.yr$^{-1}$ (55 times lower than our results obtained for iron) . The electron creation induced by irradiation facilitates the surface reactions and consequently the oxidation process. This phenomenon is also observed for example in catalytic reactions under UV exposure [4].

Figures 6 shows that the hydrogen enrichment close to 10 at.% observed for humid air and humid nitrogen atmosphere irradiations, is thus related to the presence of $H_2O$ content atmosphere. Wayne-Siek et al. [5] who measured the charged species created in an anoxic atmosphere under irradiation, have shown that more than 88% of these species are $H^+(H_2O)n$ clusters. It can be assumed that these clusters decompose at the iron surface allowing the hydrogen penetration which could explain the previous results.

4. Conclusion

It has been shown that under proton irradiation, the iron oxidation process requires the coupled action of $O_2$ and $H_2O$ in atmospheric media. In contrast, the hydrogen surface enrichment which can reach 10 at. % occurs without the presence of oxygen in humid atmosphere.

Acknowledgements

The authors would like to thank Alain Chevarier and Noëlle Chevarier for their very helpful contribution in the frame of fruitful discussions.

Figure captions

Figure 1: Irradiation set up

Figure 2: Air humidity control set up

Figure 3: Comparison of RBS and ERDA experimental spectra corresponding to the initial sample (dark curves) and the corroded one in humid air (light curves)

Figure 4: $^{15}$N profile for the 45 min iron irradiated sample in $^{15}$N (RH=45%) atmosphere

Figure 5: Comparison of oxygen concentration profiles obtained after irradiation in various gaseous media

Figure 6: Comparison of hydrogen concentration profiles obtained after irradiation in various gaseous media

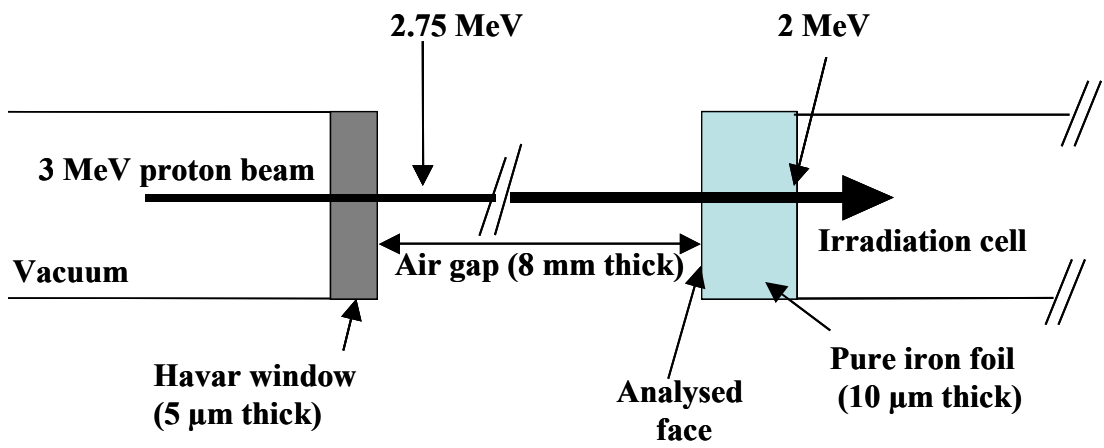

Figure 1

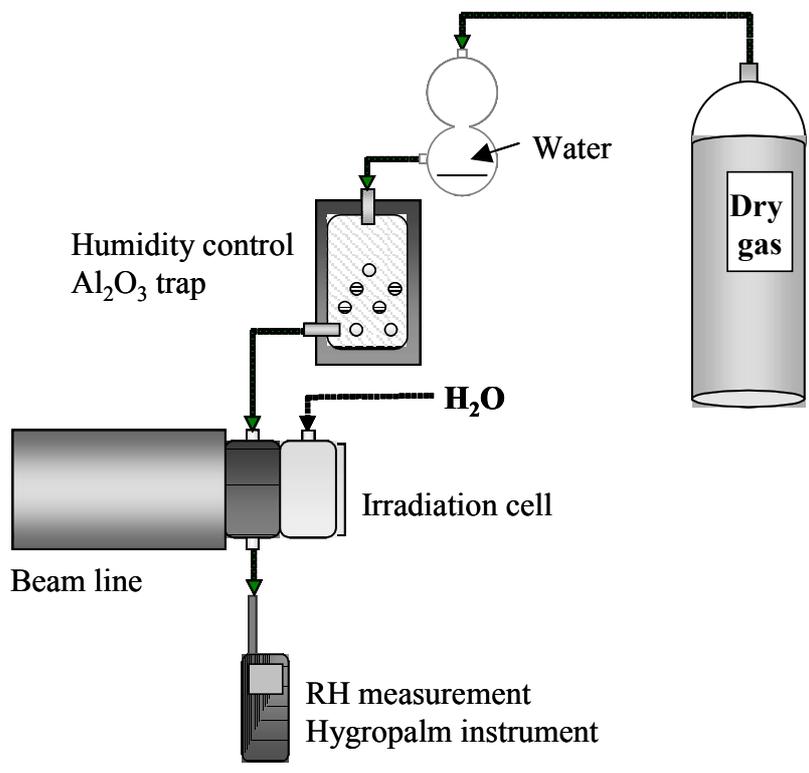

Figure 2

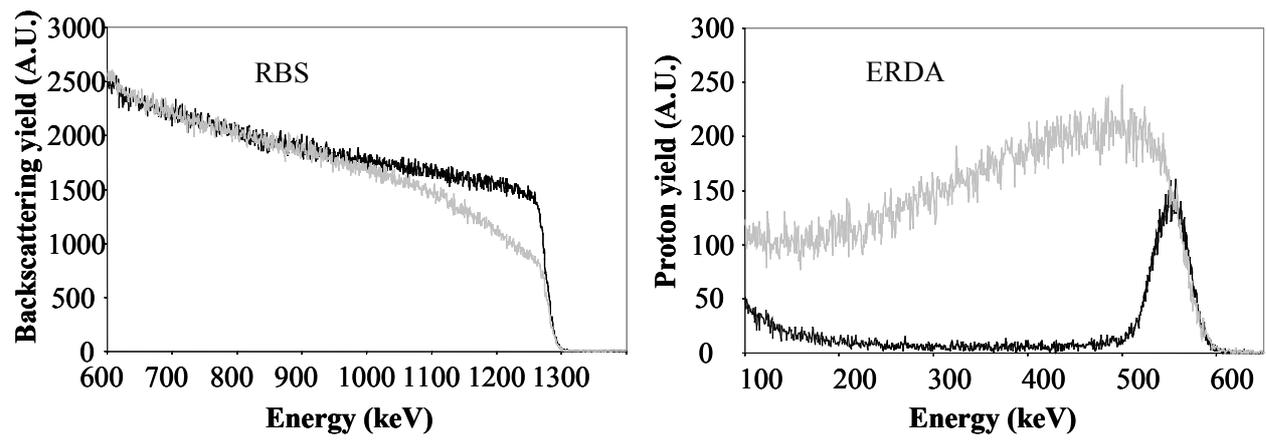

Figure 3

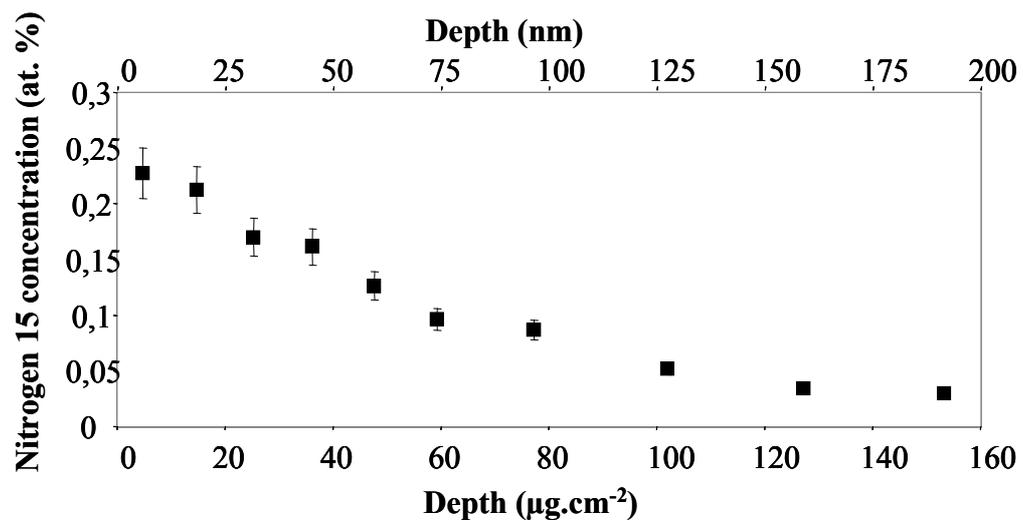

Figure 4

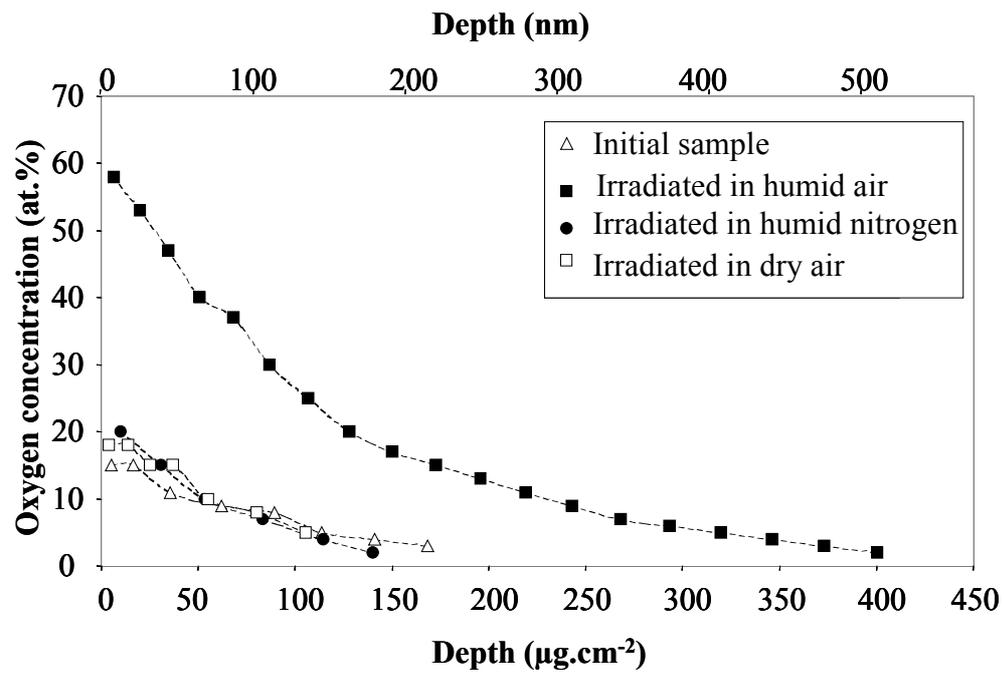

Figure 5

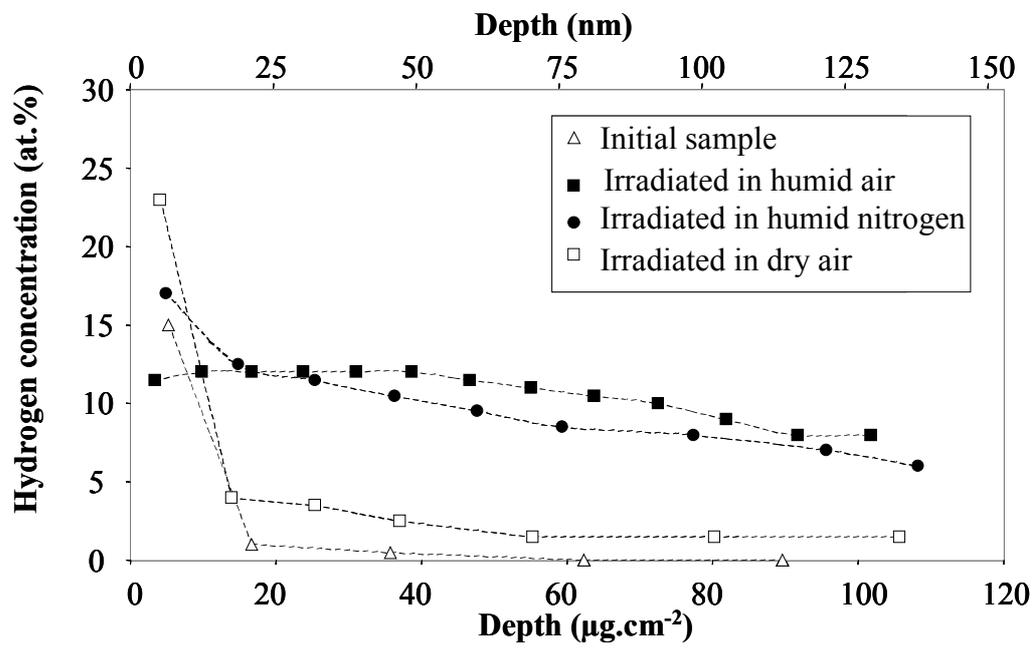

Figure 6